# Checkerboard order state in superconducting FeSe/SrTiO$_3$(001) monolayer

Cheng-Long Xue, Qian-Qian Yuan, Yong-Jie Xu, Qi-Yuan Li, Li-Guo Dou, Zhen-Yu Jia, Shao-Chun Li*

*1. National Laboratory of Solid State Microstructures, School of Physics, Nanjing University, Nanjing 210093, China*

*2. Collaborative Innovation Center of Advanced Microstructures, Nanjing University, Nanjing 210093, China*

*3. Jiangsu Provincial Key Laboratory for Nanotechnology, Nanjing University, Nanjing 210093, China*

* e-mail: scli@nju.edu.cn

**Abstract**

Ordered electronic states have been extensively explored in cuprates and iron-based unconventional superconductors, but seldom observed in the epitaxial FeSe/SrTiO$_3$(001) monolayer (FeSe/STO) with an enhanced superconducting transition temperature (Tc). Here, by using scanning tunneling microscopy/spectroscopy (STM/STS), we reveal a checkerboard charge order in the epitaxial FeSe/STO monolayer, with a period of four times the inter-Fe-atom distance along two perpendicular directions of the Fe lattice. This ordered state is uniquely present in the superconducting FeSe/STO monolayer, even at liquid nitrogen temperature, but absent in the non-superconducting FeSe monolayer or bilayer. Quasiparticle interference (QPI) measurements further confirm it as a static order without an energy-dependent dispersion and gapped out within the superconductivity gap. The intensity of the charge order shows an enhancement near the superconducting transition temperature, thus implying a correlation with the high-Tc superconductivity in the FeSe/STO monolayer. This study provides a new basis for exploring the ordered electronic states and their interplay with high-Tc superconductivity in the FeSe monolayer.

## I. INTRODUCTION

Unconventional superconductors such as cuprates and iron-based pnictides/chalcogenides usually exhibit versatile ordered states dominated by charge or spin [1-6]. There has been growing evidence that charge and spin ordered states are inherent in the high-Tc superconductors [1,2]. The checkerboard charge order was observed in $Bi_2Sr_2CaCu_2O_{8+\delta}$ and $Ca_{2-x}Na_xCuO_2Cl_2$ [2,3]. It was subsequently found that the charge ordered states cover a wide range of the phase diagram of cuprates, although their nature are heavily debatable [7]. The iron-based superconductor family is another platform for exploring the electronic ordered states [8-11], in which the superconducting state is usually located in the proximity of a magnetically ordered state [9,11] and the electronic nematic order is a ubiquitous feature [12-18]. Nuclear magnetic/quadrupole resonance measurements showed the charge order state in $AFe_2As_2$ (A = Rb, Cs) [19,20], and STM measurements showed the charge order in doped $BaFe_2As_2$ [21]. It was reported that strain can induce stripe-like charge order in LiFeAs that simultaneously suppresses the superconductivity [22]. A checkerboard order in the iron lattice was also found in bulk FeSe [23].

Bulk FeSe exhibits a superconducting transition (Tc) at ~9 K [24], while the epitaxial monolayer of FeSe on $SrTiO_3(001)$ substrates (FeSe/STO) exhibits a surprisingly enhanced Tc [25-41]. The interfacial charge transfer can shift downward the hole pocket at the Brillouin zone (BZ) center, and result in an Fermi surface consisted of only electron pockets at the BZ corner [42-44]. In FeSe/STO bilayer and multilayer, an incommensurate smectic or a stripped nematicity electronic state was observed [12,17,45]. However, no electronic ordered state has been reported in the superconducting FeSe/STO monolayer to date. It's noteworthy that even though sharing an analogous Fermi surface, the $K_xFe_{2-y}Se_2$ exhibits a 2×2 and 2×$\sqrt{10}$ charge order in superconducting bulk and epitaxial thin films, respectively [46,47].

Scanning tunneling microscopy/spectroscopy (STM/STS) in combination with quasiparticle interference (QPI) technique is a powerful tool in revealing the real-space electronic structures, and has been widely applied to explore the electronic ordered states in high-$T_c$ superconductors [2-4,7,14,15,45-47]. By using STM/STS-QPI technique, we discovered a unique static charge order state in FeSe/STO monolayer, with a period of four times the inter-Fe-atom distance along two perpendicular directions of the Fe lattice. The intensity of the checkerboard order shows an enhancement near ~ 40 K. Surprisingly, this checkerboard ordered state is absent in the non-superconducting FeSe monolayer as considered as the parent phase in previous studies [48], and the FeSe bilayer, thus implying a correlation between the checkerboard state and the high-Tc superconductivity in the FeSe/STO monolayer.

## II. EXPERIMENT METHODS

The FeSe monolayer and bilayer were epitaxially grown on the SrTiO$_3$ (001) substrate by using molecular beam epitaxy (MBE) technique in an ultrahigh vacuum chamber (The base pressure is less than 1×1$^{-10}$ Torr). The SrTiO$_3$ (001) substrates were loaded into ultrahigh vacuum and degassed at ~600 °C for 3 hours, annealed at ~1000 °C for 30 min, and subsequently annealed at ~1250 °C for 20 min to obtain a TiO$_2$-terminated surface. High purity Fe (99.995%) and Se (99.9999%) were evaporated from two standard Knudsen cells, and the SrTiO$_3$ substrates were kept at ~420 °C during growth. The superconducting FeSe/STO monolayer was obtained through annealing the as-grown sample up to ~530 °C for hours. STM experiments were *in-situ* performed with a combined low temperature

STM-MBE system (Unisoku, USM1500) at ~4.2 K. A polycrystalline Pt-Ir tip was used for scan. The topographic images were taken under the constant current mode, and the dI/dV spectra (STS) were collected using a standard lock-in technique with a bias modulation at 879 Hz.

## III. RESULTS AND DISCUSSIONS

Figure 1(a) illustrates the atomic structure of FeSe/STO monolayer. The as-grown FeSe/STO monolayer initially exhibits no superconductivity transition, and a post annealing can induce the high-Tc superconductivity transition, as illustrated in Fig. 1(b). Figures 1(c)-1(j) show the STM and STS results taken on the as-grown and superconducting FeSe monolayer (see also supplementary Fig. S1 [49]). The surface of as-grown monolayer FeSe, Fig. 1(c), is decorated by bright Se-like clusters [36,50]. The STS spectrum in Fig. 1(g). According to the previous studies, we believe the as-grown sample is not superconducting even though having a gap-like feature. (We attempted to fit the as-grown spectrum with a Dynes function, assuming a superconducting gap. However, the fitting result is not accurate at all, and can only be kept for either the coherence peaks or the gap depth). Upon annealing to ~450°C, the Se-like clusters are removed from the surface, and some bright dumbbell-like pairs start to appear [Fig. 1(d)], which can be attributed to Fe vacancies [51-53]. Meanwhile, the FeSe monolayer starts to undergo a superconductivity transition, as featured by the emergence of a superconducting energy gap in the STS spectra, as shown in Fig. 1(h). Unexpectedly, nanoscale patches of a checkerboard pattern starts to appear simultaneously with a periodicity of four times the inter-Fe-atom distance along two perpendicular directions, as shown in Fig.1(d). After further annealing to enhance the superconductivity energy gap, the checkerboard pattern becomes more prominent, as shown in Figs. 1(e) and 1(i). To better envision the checkerboard pattern, we optimized the quality of FeSe monolayer and obtained the largest superconducting energy gap, as shown in Figs.1(f) and 1(j). The checkerboard pattern is found to cover the whole FeSe surface, while slightly suppressed in the vicinity of surface defects (STM/STS data can be found in Supplementary Figs. S2 and S3 [49]). It is noteworthy that the checkerboard pattern is absent in the as-grown non-superconducting FeSe/STO monolayer.

To further examine whether the checkerboard pattern is uniquely present in the superconducting state, we also looked at the FeSe/STO bilayer. Different from the FeSe/STO monolayer, the second layer of FeSe is not superconductive [8,25,50]. Surprisingly, no such checkerboard pattern is observed in the FeSe bilayer (STM data can be found in Supplementary Fig. S4 [49]). This is different from the recently reported smectic/nematic orders that are only present in FeSe bilayer/multilayers [12,17,45].

Differential conductance dI/dV maps were measured on the superconducting FeSe monolayer under various bias voltages, as shown in Fig. 2(a)-2(c). The fast Fourier transform (FFT) images of the dI/dV maps, as represented in Figs. 2(d)-2(f), are used to determine the energy-dependent wave vector of this checkerboard order (see also Supplementary Fig. S3 [49]). The FFT results near Fermi energy, Fig. 2(f), as characterized by the ring-like QPI features, reproduce very well the previous studies [30,52,54]. As the bias voltage increases, for instance, Figs. 2(d) and 2(e), scattering vectors associated with the checkerboard state becomes strikingly visible. The wave vectors extracted from these FFT images are plotted in Fig. 2(g). It is confirmed that the checkerboard state is a static electronic order, without a bias-dependent dispersion (see also Supplementary Fig. S3 [49]). However, the charge order is nearly invisible around $E_F$, as shown in Figs. 2(f) and Supplementary Fig. S3 [49], particularly in the region within ~ ±10 mV, which seems to be suppressed by the superconducting energy gap.

To investigate the impact of checkerboard state to the spatial variation of electronic structures, we have measured the STS spectra along the high symmetry direction. Figure 3(a) shows a highly-resolved STM image, and Fig. 3(b), the dI/dV spectra collected along the black arrowed line marked in Fig. 3(a). A weak but discernible spatial modulation can be identified in the spectral intensity, especially at the positive bias voltage. In Fig. 3(c) is plotted the spectral intensity extracted near the coherence peaks as a function of the position. The intensity modulation clearly follows the period of the checkerboard order. Moreover, the spectral intensity near the first and second coherence peaks show an antiphase modulation, as shown in Fig. 3(c). According to the previous studies [55,56], the superconducting gap shows an anisotropy on the electron pockets, and is related with the multi-orbital physics. Thus the anti-correlated modulation of the two coherent peaks, as shown in Fig. 3(c), is possibly due to the responses of different orbitals to the checkerboard period as well. Besides, the superconducting gap size, as defined by the half of distance between the two largest coherence peaks, was also extracted as a function of the position. As plotted in Fig. 3(e), the superconducting gap size is also found to be spatially modulated, consistent with the period of the checkerboard order. These results thus reveal the correlation between the checkerboard order and superconductivity.

We performed the temperature-dependent measurements, to further unveil the relationship between the checkerboard state and superconductivity. The checkerboard state is observable within the whole temperature range explored (from ~4 K to ~77 K). Figures 4(a)- 4(d) show the select STM images taken at various temperatures (see also Supplementary Fig. S5 [49]). In Fig. 4(e) is plotted the intensity of the checkerboard order versus temperature, as defined by the ratio of its FFT peaks to the lattice Bragg ones. Upon cooling down from 77 K to 4 K, the intensity of the checkerboard state shows a fast rise followed by a monotonous drop, with a peak located at around ~40 K. This peak is coincident with the Tc of ~40 K in the FeSe/STO monolayer as determined by the electrical transport measurements in previous studies [27,57,58]. A similar enhancement of the charge order near Tc has also been observed in (Y, Nb)Ba$_2$Cu$_3$O$_{6+x}$, implying a competition between the charge order and the superconductivity [6].

Now we turn to discuss the origin of the checkerboard ordered state. It is well known that the interplay between Fe vacancy, magnetism and superconductivity plays an essential role in iron-based superconductors. First of all, we rule out the possibility of a Fe vacancy ordered state. The Fe vacancy order has been considered as the parent compound for FeSe-related superconductors [59-61]. However, all of these Fe vacancy ordered states are insulating, in contrast to the checkerboard ordered state as observed in the superconducting FeSe/STO monolayer. Moreover, the checkerboard state are very different from a well-defined Fe vacancy in topographic and spectroscopic characterizations (see Supplementary Figs. S2 and S6 [49]).

The strong bias-dependence of the checkerboard pattern points to the indication of its electronic origin. However, the case that the checkerboard order is only formed in the superconducting FeSe monolayer is in contrast with the previously reported smectic/nematic orders in FeSe bilayer or multilayers [17,45]. Such checkerboard state doesn't break the C$_4$ symmetry, although breaking the lattice symmetry, suggesting that it is not the nematic/smectic origin. Considering the checkerboard pattern modulated superconducting coherence peaks, Figs. 3(b) and 3(c), as well as the temporal evolution of the checkerboard intensity, Fig. 4(e), our checkerboard order is phenomenologically analogous with the electronic ordered state in cuprates and iron-based superconductors [6,7,46].

Since a nonmagnetic tip was used in our measurements, the electronic information was provided mainly without a spin resolution. However, the mechanism of Fermi surface nesting regardless of spin cannot simply realize the period of the checkerboard order. Moreover, even though the Fermi surface can be significantly varied upon annealing [48], the period of the checkerboard state is always kept unchanged. If there exists a weak coupling between the spin and charge scattering channels, it is also possible that the checkerboard pattern reflects the magnetic orders of FeSe monolayer as well [62]. In the alkali-doped FeSe superconductors, a charge-density modulation is caused by the block-antiferromagnetic ordering of the iron moments [46]. Even though the magnetic order of FeSe/STO monolayer has not been pinned down, density functional theory (DFT) calculation reproduces very well the experimental results if assuming a checkerboard antiferromagnetic order [63-65].

The previous studies reported the 2×1, is originated from the electronic periodicity caused by the STO surface superstructure [50,66,67]. Other periodicities such as 3×1, $\sqrt{2}\times\sqrt{2}$, $\sqrt{13}\times\sqrt{13}$ and 4×1, are also due to the substrate-induced surface reconstruction [68-70]. For the checkerboard order as observed in the superconducting FeSe/STO monolayer, even though having a similar interface, the non-superconducting FeSe monolayer exhibits no checkerboard order, as shown in Fig.1(c). Thus, the possibility that the checkerboard order comes from the direct projection of substrate's surface/interface atomic reconstruction can be excluded. Even so, we believe the formation of the checkerboard is still associated with the STO substrate, considering that its intensity shows an increase near Tc that is enhanced by the substrate. It is possible that the substrate impacts the checkerboard order formation in a similar way to the interface-enhanced superconductivity. It has been reported that interface could enhance the antiferromagnetic interaction, and helps to maintain the large spin fluctuations under heavy electron doping [63-65,71]. The nematicity only emerged in the FeSe/STO multilayer with a decreasing tensile strain and suppressed charge transfer [8,12,17], and the smectic state in FeSe/STO bilayer may be originated from a balance between strain and charge [45]. Similarly, in the FeSe/STO monolayer, the interfacial strain, charge transfer, spin fluctuation and enhanced electron-phonon coupling may together enhance the electronic correlation and give rise to the formation of the checkerboard order.

## IV. SUMMARY

In summary, we discovered a new checkerboard charge order in the superconducting FeSe/STO monolayer. Although the origin is still unclear, the checkerboard order shows a strong correlation with the superconductivity. Our results lay a foundation for further investigations to probe the nature of the ordered states and their correlation with superconductivity in high-Tc iron-based superconductors.

## ACKNOWLEDGMENTS

This work was financially supported by the National Key Research and Development Program of China (Grants No. 2021YFA1400403), the National Natural Science Foundation of China (Grants No. 92165205, No. 11790311, No. 11774149) and Innovation Program for Quantum Science and Technology (Grant No. 2021ZD0302800).

**Figures and captions**

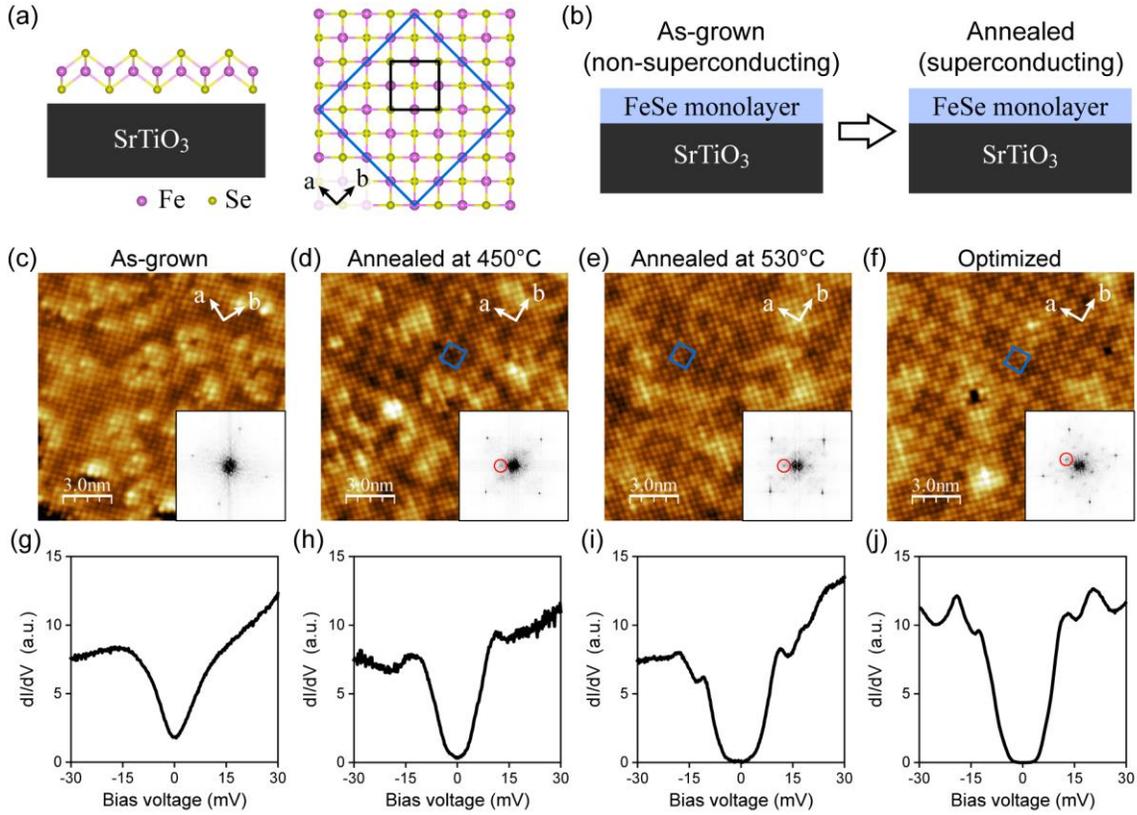

FIG. 1. Formation of the checkerboard pattern in superconducting FeSe/STO monolayer. (a) Atomic structure (left: side view; right: top view) of the epitaxial FeSe grown on SrTiO$_3$ substrate. The black and blue squares mark the normal unit cell of top Se atoms and the checkerboard order, respectively. The a and b axes are defined by the Fe lattice. (b) Schematic illustration of the annealing-induced superconducting transition in the FeSe/STO monolayer. (c)-(f) Atomically-resolved topographic images (15×15 nm$^2$) of the as-grown and superconducting FeSe/STO monolayer. $U = +100$ mV and $I_t = 1$ nA. Blue squares in (d)-(f) outline the unit cell of the checkerboard patterns. The insets are the corresponding FFT images, and the red circles mark the vectors of the checkerboard charge order in (d)-(f). (g)-(j) dI/dV spectra taken on the surfaces of (c)-(f), respectively. Set point: $U = +30$ mV and $I_t = 1$ nA.

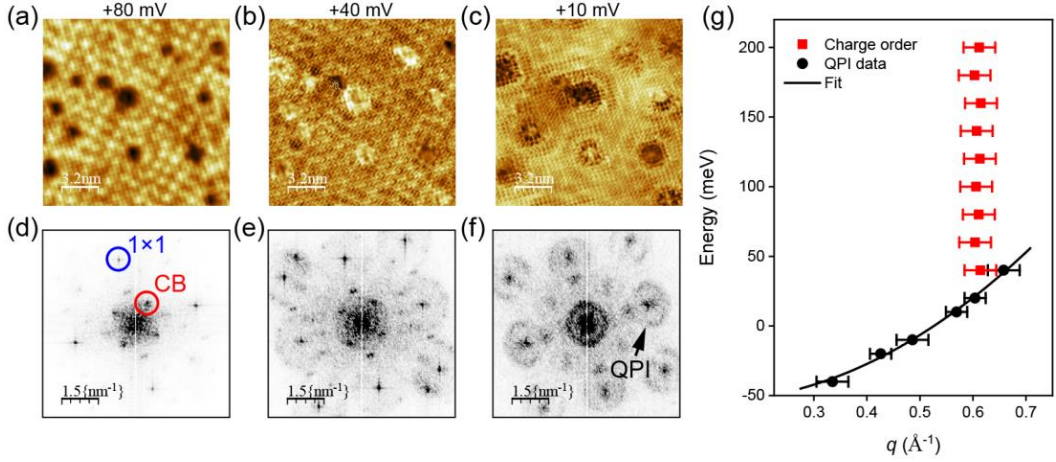

FIG. 2. Differential conductance maps of the superconducting FeSe/STO monolayer. (a)-(c) Differential conductance maps taken on the superconducting FeSe/STO monolayer. The corresponding bias voltage is labeled above each image. (d)-(f) Fast Fourier transform (FFT) images of the dI/dV maps as shown in (a)-(c), respectively. The red and blue circles marked in (d) are the vectors of the checkerboard periodicity and the lattice Bragg points, respectively. The ring-like features marked by the black arrowed line in (f) are originated from the quasiparticle interference (QPI) of electrons. (g) Energy dependence of the wave vectors extracted from the FFT of dI/dV maps taken at various bias voltages. The red dots represent the wave vectors from the checkerboard periodicity, and the black ones from the QPI features. The black curves are a parabolic fitting to the dispersive QPI wave vectors.

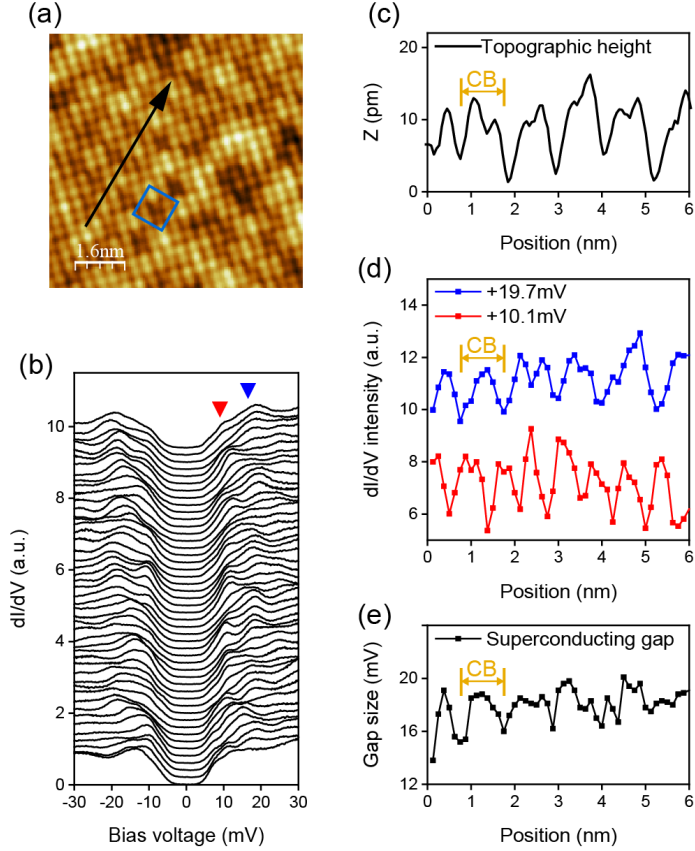

FIG. 3. Spatial modulation of the dI/dV spectra by the checkerboard pattern. (a) Atomically-resolved topographic image ($8 \times 8 nm^2$) showing the checkerboard pattern in superconducting FeSe/STO monolayer. $U = +100$ mV, $I_t = 200$ pA. The blue square marks the unit cells of the checkerboard periodicity. (b) A series of dI/dV spectra taken along the black arrowed line in (a). The spectra are uniformly shifted in the vertical direction for clarity. The red and blue triangles mark the position of the first and second superconducting coherence peaks, respectively. Set point: $U = +30$ mV, $I_t = 1$ nA. (c) Line-scan profile taken on the black arrowed line in (a). The $x$ and $y$ axis represent the length of the measured line-scan profile and STM apparent height, respectively. (d) dI/dV spectral intensities extracted from the positions as marked by the blue and red triangles in (b). (e) The superconducting gap size extracted from the dI/dV spectra in (b). The $x$ axis in (d) and (e) represents the real space positions along the black arrowed line in (a).

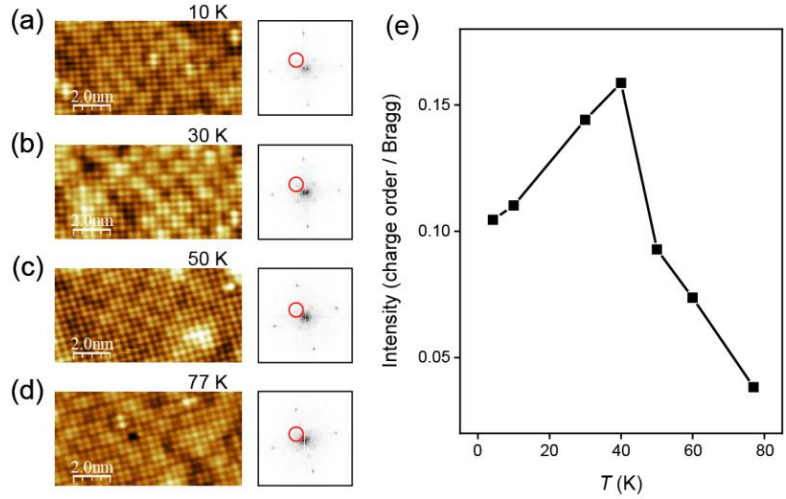

FIG. 4. Temperature dependence of the checkerboard pattern. (a)-(d) Atomically-resolved topographic images (10 × 5 nm$^2$) obtained at various temperatures as labelled above each image. All the images were collected under the same tunneling condition. $U = +100$ mV, $I_t = 1$ nA. The insets are the corresponding FFT images, and the red circles mark the vectors of the checkerboard charge order. (e) Temperature dependence of the intensity of the checkerboard pattern. The intensity of the checkerboard state is defined by the ratio of its FFT peaks to the lattice Bragg ones.

(2018).